# MAPUNetR: A Hybrid Vision Transformer and U-Net Architecture for Efficient and Interpretable Medical Image Segmentation.

Ovais Iqbal Shah[1], Danish Raza Rizvi[2] Aqib Nazir Mir[3*]

[1,2,3]Department of Computer Engineering Jamia Millia Islamia (A Central University) New Delhi India 110025

Email:[1]dk90910@gmail.com, [2]drizvi@jmi.ac.in, [3]aqib2206142@st.jmi.ac.in

[*]Corresponding Author

## Abstract

Medical image segmentation is pivotal in healthcare, enhancing diagnostic accuracy, informing treatment strategies, and tracking disease progression. This process allows clinicians to extract critical information from visual data, enabling personalized patient care. However, developing neural networks for segmentation remains challenging, especially when preserving image resolution, which is essential in detecting subtle details that influence diagnoses. Moreover, the lack of transparency in these deep learning models has slowed their adoption in clinical practice. Efforts in model interpretability are increasingly focused on making these models' decision-making processes more transparent. In this paper, we introduce MAPUNetR, a novel architecture that synergizes the strengths of transformer models with the proven U-Net framework for medical image segmentation. Our model addresses the resolution preservation challenge and incorporates attention maps highlighting segmented regions, increasing accuracy and interpretability. Evaluated on the BraTS 2020 dataset, MAPUNetR achieved a dice score of 0.88 and a dice coefficient of 0.92 on the ISIC 2018 dataset. Our experiments show that the model maintains stable performance and potential as a powerful tool for medical image segmentation in clinical practice.

## 1. Introduction

Medical image segmentation has become an indispensable technique across various biomedical research and clinical practice domains, offering immense utility and numerous opportunities for scientific advancement. Using this essential method, researchers can precisely mark the boundaries between anatomical features and areas of pathology, facilitating applications such as brain tumour segmentation, breast tumour detection, and lesion characterization. By utilizing advanced segmentation algorithms, medical professionals can accurately define areas of interest, providing essential visual data for diagnostics and treatment planning. The advent of data-driven approaches, particularly artificial intelligence(A.I.), has significantly accelerated progress in medical image segmentation. AI-powered segmentation methods now outperform conventional techniques, with research increasingly focused on exploring the "black-box" nature of these models to improve transparency and trust in clinical settings. The demand for interpretable and explainable AI-driven segmentation models has emerged as a critical research priority, aiming to enhance their adoption in real-world medical workflows. In medical image segmentation, accuracy is paramount. Errors that may be acceptable in natural image processing can have severe consequences in clinical applications, where even the slightest misclassification can lead to incorrect diagnoses or suboptimal treatment plans [1].

There has been a fundamental transformation in how medical images are processed and segmented. Once dominated by convolutional neural networks (CNNs), has now transitioned towards transformer-driven architectures, demonstrating superior capabilities in capturing complex, long-range dependencies within images. Such computational frameworks initially developed to process natural language processing(NLP) tasks, have proven to be highly effective in medical imaging, where understanding intricate spatial relationships between regions is essential. The traditional fully convolutional neural networks (FCNNs) with U-shaped encoder-decoder structures like U-Net architecture [2], were previously the gold standard in segmentation tasks. U-Net's encoder progressively down samples feature representations to capture global context, while the decoder up samples to recover fine-grained details, facilitating pixel-wise or voxel-wise segmentation. This architecture has demonstrated strong performance in diverse segmentation applications, but the rise of transformers offers new advantages. Transformers, unlike CNNs, do not rely on convolutions but relies on self-attention patterns to capture relationships between all parts of the image. This shift enables them to capture more comprehensive and global patterns between different patches of the image, a critical requirement in medical segmentation tasks where precision and detail are vital. As a result, transformers have demonstrated remarkable success in addressing the challenges posed by medical imaging, including segmenting highly variable and complex structures [3]. Their versatility across various applications, from text processing to image analysis, highlights the adaptability and potential, as depicted in Figure 1.

AI has already demonstrated its ability to match or surpass human expertise in various medical specialities such as cardiology, dermatology, ophthalmology, radiology, and pathology. These A.I. systems enhance diagnostic accuracy and streamline the analysis process, making them indispensable tools in modern healthcare [4]. AI's ability to efficiently analyze large volumes of visual data is especially valuable in medical imaging, enabling faster diagnosis while preserving high levels of accuracy [5].

To address the gaps in existing medical image segmentation models and further enhance the capabilities of hybrid architectures, this research proposes several key questions. These questions aim to explore the potential of integrating Vision Transformers with U-Net architectures, focusing on segmentation performance, model interpretability, and real-world applicability in clinical environments:
1. How does incorporating Vision Transformer elements with U-Net frameworks lead to better segmentation results compared to traditional convolutional models across various medical imaging datasets, such as BraTS 2020 and ISIC 2018?
2. What role do attention mechanisms in Vision Transformers play in improving the model's strength in processing both overall image relationships and fine spatial features in medical segmentation tasks?
3. How can the interpretability of transformer-based models be further enhanced to offer clearer insights into the decision-making process, addressing the critical need for explainable AI in clinical applications?
4. How well does this new hybrid approach perform relative to other modern transformer architectures on medical image segmentation benchmarks in terms of accuracy, generalizability, and computational efficiency?

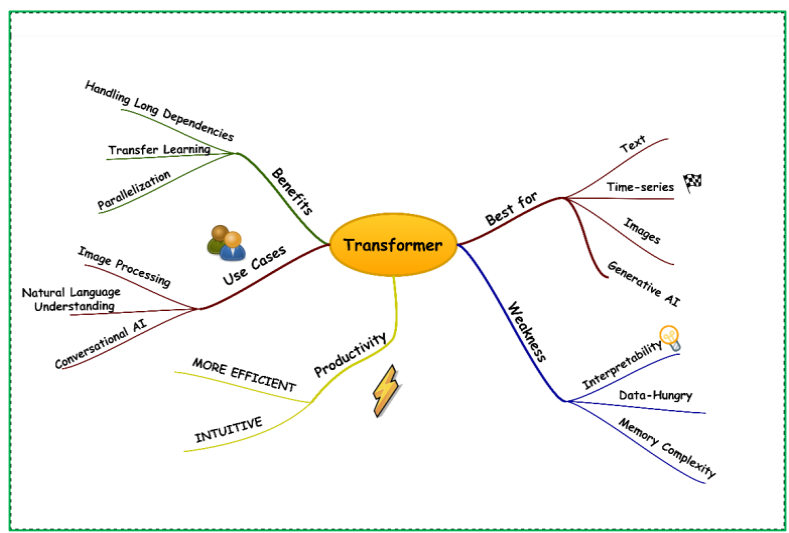

Figure 1:Demonstration of Transformers.

This paper is structured as follows: Section 2 presents a comprehensive literature survey, exploring existing knowledge in-depth and identifying research gaps. Following this, Section 3 delves into the materials and methods, elucidating the techniques and methodologies employed in the research. Section 4 is dedicated to presenting the experiments and results, accompanied by a comprehensive discussion of the findings. Finally, Section 5 provides a conclusion, discussing key findings and their implications.

## 2. Related Works
Over the past half-decade, remarkable progress has been achieved in the field of medical image segmentation, largely driven by the introduction of transformer-based neural networks. The transformation in architectural design has dramatically improved how we perform segmentation, enabling more accurate identification and mapping of anatomical components in medical images. This structural approach proves especially powerful because it can process both distant relationships and comprehensive contextual information, making it ideal for detailed medical segmentation work. As a result, these models have become the focal point of recent research efforts to enhance segmentation accuracy and model interpretability.

One noteworthy advancement in this domain is presented by Farrag et al.[6] who introduced an innovative dual-dilated convolution module designed to address the challenge of balancing local spatial detail with global contextual understanding. The dual-dilated convolution approach was engineered to preserve fine-grained spatial details while expanding the receptive field, enabling the neural network to capture both local and global context effectively. This capability is particularly important for medical image analysis, as exact spatial accuracy is necessary to properly define and map pathological regions. Farrag et al. conducted rigorous evaluations of their approach using Gradient-weighted Class Activation Mapping (Grad-CAM) to assess the interpretability of the model alongside comprehensive quantitative comparisons with other explainable segmentation techniques. Their work highlights the importance of explainability in clinical applications, where trust in A.I. systems is paramount.

In another pivotal study, Heidari et al. [7] introduced HiFormer, which strategically integrates CNN architecture with transformer technology to achieve improved medical image segmentation results. The HiFormer architecture synergistically combines the strengths of both paradigms, using a double-level fusion (DLF) module strategically integrated within the skip connections of the encoder-decoder architecture. The DLF module effectively fuses multi-scale features, combining high-level global context with low-level spatial details, thereby improving segmentation accuracy. This architectural innovation represents a significant leap in addressing the limitations of pure CNN or Transformer models, particularly in medical applications where global context and fine detail are critical for accurate segmentation.

Building on this, the hierarchical Swin transformer based TransDeepLab architecture [8], proposed by another group of researchers, further advances the field by integrating shifted windows into the DeepLabv3 framework [9]. This architecture enhances the atrous spatial pyramid pooling (ASPP) module, enabling the model to capture multi-scale contextual information more effectively. The TransDeepLab model employs successive Swin transformer blocks interleaved with patch-merging operations, substantially reducing spatial detail while maintaining high-level feature extraction. The architectural design of TransDeepLab has demonstrated superior performance, achieving an impressive dice similarity coefficient (DSC) of 80.16% and a Hausdorff distance (H.D.) of 21.25%, established new performance standards in the field of medical image segmentation.

In another work on U-shaped networks, the authors of BEFUnet [10] introduced a design that enhances medical image segmentation by simultaneously processing edge details and body features. The key innovation of BEFUnet lies in the local cross-attention feature (LCAF) fusion module, which employs attention mechanisms to combine features from different parts of the network, thus enhancing spatial relationships. This is complemented by a double-level fusion (DLF) module that merges low-level and high-level features, ensuring that detailed local information and broader context are preserved. BEFUnet's dual-branch encoder, which processes body and edge information in parallel, further enhances segmentation by enabling the model to handle different aspects of the image separately before integrating them. Across diverse medical imaging scenarios, this architectural approach achieved outstanding segmentation results, including a DSC of 80.47% and H.D. of 16.26% on the Synapse dataset [11] and an accuracy of 95.16% on the SegPC-2021 dataset [12]. Additionally, BEFUnet exhibited exceptional performance on the ISIC 2017 dataset [13], achieving a DSC of 86.8%, sensitivity (S.E.) of 85.3%, specificity (S.P.) of 98.5%, accuracy (ACC) of 94.6%, and an intersection over union (IoU) of 76.8%.

Aboussaleh et al. [14] engineered an advanced 3D hybrid solution that combines multiple MRI data types to achieve precise brain tumor segmentation. Their model, assessed on the BraTS 2020 dataset [15] [16] [17], achieved highly competitive DSCs of 91.95% for whole tumor (W.T.), 82.80% for tumor core (T.C.), and 81.70% for enhancing tumor (E.T.), underscoring its efficacy in segmenting complex brain tumour subregions. Using multi-modal data and integrating advanced 3D segmentation techniques allow for a more nuanced analysis of brain tumours, further pushing the boundaries of medical segmentation technology.

Lastly, a more recent study, a novel 3D Transformer-based architecture, Swin UNETR, was introduced [18]. This model integrates a Swin Transformer [19] as the encoder with a CNN-based decoder, utilizing hierarchical encodings and skip connections to process and analyze 3D medical images effectively. The Swin UNETR architecture is pre-trained and self-supervised, incorporating auxiliary tasks to improve the model's effectiveness at identifying and mapping different areas within medical scans. The hierarchical structure of the encoder, combined with the power of transformers, enables the implementation of Swin UNETR led to breakthrough performance in dividing medical images into distinct regions, offering a highly effective solution for complex segmentation tasks.

## 3. Materials and Methods

The proposed approach involves multiple stages, beginning with data collection, followed by rigorous pre-processing techniques, and culminating in patchification for the Vision Transformer model. Each component in this pipeline plays a crucial role in preparing the medical images for effective segmentation. The detailed methodology workflow is given in Figure 2.

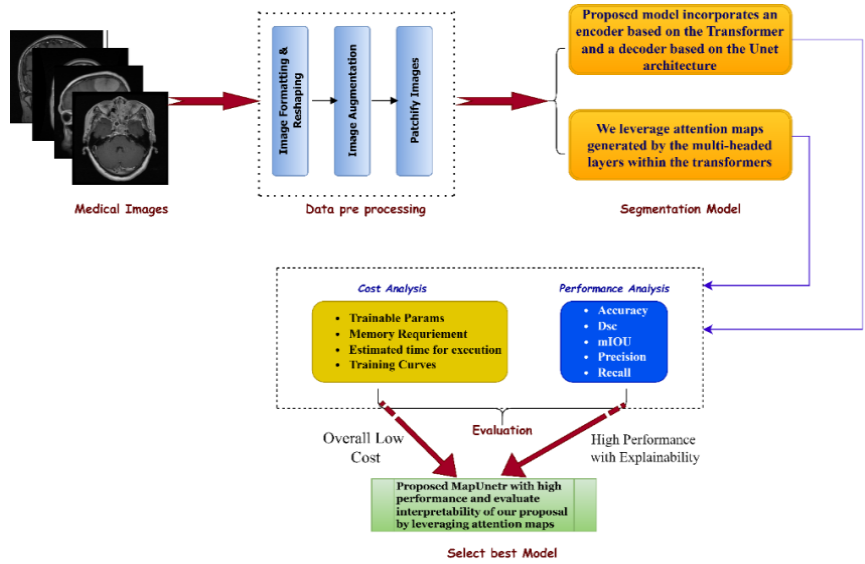

Figure 2: General overview of the workflow of the proposed methodology for image segmentation.

### 3.1 Data Collection and Exploratory Data Analysis (EDA)

The dataset $D$ consists of pairs $(I, M)$, where $I$ represents an image and $M$ is the corresponding segmentation mask. The collection and analysis of data involve identifying important visual patterns and checking for inconsistencies. We conduct statistical and visual analysis in the exploratory phase to ensure the data conforms to the model's input requirements. Common tasks include calculating pixel intensity distributions, mask-to-image correspondence checks, and identifying anomalies such as incomplete or misaligned masks.

### 3.2 Data Pre-processing

Effective data pre-processing is crucial for maintaining the quality and consistency of medical images used in segmentation tasks. This section outlines the pre-processing techniques applied to the datasets, including image formatting, augmentation, normalization, and partitioning.

### 3.2.1 Image Formatting

Each image $I$ in the dataset is represented as $I \in \mathbb{R}^{H \times W \times C}$, where $H$ and $W$ denote the height of the input image and width of input image respectively, and $C$ represents the number of channels. For consistency, the images and masks are resized to a fixed dimension $R \times R$ times, ensuring uniform input sizes across the dataset:

$$I_{formatted} = rsize(I, R \times R), \quad M_{formatted} = rsize(M, R \times R) \qquad (1)$$

This resizing ensures that the model processes images of a standard dimension, reducing the risk of inconsistencies during training and evaluation.

### 3.2.2 Image Cropping

In medical imaging, large portions of an image may consist of irrelevant background, which can increase computational complexity and introduce biases during training. For the BraTS2D dataset, where regions of interest occupy smaller portions of the image, cropping the images to a smaller dimension $128 \times 128 \times 3$ was necessary. This reduction focuses the model on the relevant anatomical structures. The cropping function $crop\ (I, d)$ trims the image to a specified dimension $d$, ensuring the model primarily learns from the most informative parts:

$$I_{cropped} = crop(I, 128 \times 128 \times 3) \tag{2}$$

Cropping not only reduces computational complexity but also minimizes model overfitting by removing irrelevant background noise.

### 3.2.3 Image Augmentation

In medical imaging, acquiring large datasets can be challenging due to the need for expert annotations. As a result, augmentation plays a vital role in increasing the diversity of training samples without manually collecting new data. To address overfitting and improve the model's generalization, various augmentation techniques are applied to the training images. These augmentations are mathematically expressed as transformations $T$ applied to each image $I$:

$$I_T = T(I), T \in \{rotate, flip, distortion, crop\} \tag{3}$$

The following augmentation strategies were employed:
*Center crop:* Extracts a central subregion from the image to focus on key anatomical areas.
*Random rotate 90:* Rotates the image by a random multiple of 90 degrees:

$$I_T = R_{90}(I), R_{90} \in \{0°, 90°, 180°, 270°\} \tag{4}$$

*Grid distortion:* Applies spatial distortion by dividing the image into a grid and randomly altering the grid points, introducing non-linear variations in the image structure.
*Horizontal and vertical flip:* Flips the image along the horizontal or vertical axis to simulate different viewing angles.
Let $I \in \mathbb{R}^{H \times W \times C}$ represent the original image. After augmentation, the transformed image $I_T$ is derived from the original via various transformations $T$, such that:

$$I_T = T(I) = \{T_1 I, T_2 I, \ldots, T_n I\} \tag{5}$$

Where each transformation $T_i$ includes rotations, flips, and distortions. Augmentation can be expressed as an operator $T$ acting on the image space:

$$I_T = T_k(I), k \in \{1,2,3,\ldots,K\} \tag{6}$$

Where $K$ represents the number of transformations applied to the image $I$. The set $\{I_T\}$ forms the augmented dataset, which is subsequently used for training the model.
These augmentations help the model learn to recognize anatomical structures under varied orientations and conditions, thus improving generalization to unseen data.

### 3.2.4 Image Normalization

Image normalization is employed to standardize pixel intensity values across the dataset. Normalizing images to a consistent range, such as [0, 1], or standardizing based on mean and standard deviation, ensures that all inputs have similar intensity distributions, which aids in model convergence. The normalization formula is defined as:

$$I_{norm} = \frac{I-\mu}{\sigma} \tag{7}$$

Where $\mu$ and $\sigma$ represent the mean and standard deviation of the pixel intensities within the dataset. Alternatively, the pixel values can be rescaled to the [0, 1] range by:

$$I_{norm} = \frac{I-min(I)}{max(I)-min(I)} \tag{8}$$

Normalization ensures that the model receives inputs that are scaled consistently, leading to faster and more stable convergence during training.

### 3.2.5 Patchification for Vision Transformers

Given that Vision Transformers operate on sequences of tokens rather than 2D grids, it is necessary to convert the input images into a sequence of patches. Each image $I \in \mathbb{R}^{H \times W \times C}$ is divided into non-overlapping patches of size $P_h \times P_w$, which are then flattened and treated as individual tokens. The total number of patches $N$ is calculated as:

$$N = \frac{H \times W}{P_h \times P_w} \tag{9}$$

For an image of size $256 \times 256$ and a patch size of $16 \times 16$, the image is partitioned into:

$$N = \frac{256 \times 256}{16 \times 16} \tag{10}$$

Each patch $P_k \in \mathbb{R}^{P_h \times P_w \times C}$ is flattened into a vector $T_k \in \mathbb{R}^{P_h \times P_w \times C}$, where $k = 1, 2, .., N$. The image is thus transformed into a sequence of patches, which serve as input tokens for the Vision Transformer:

$$I_{patchified} = \{P_1, P_2, \ldots, P_{265}\} \tag{11}$$

This process enables the vision transformer to learn contextual relationships between different regions of the image in a manner analogous to how it processes sequences in NLP tasks.

Through data formatting, cropping, augmentation, normalization, and patchification, the images are prepared for the vision transformer based model. Each of these steps plays an essential role in optimizing the input data for effective model training, ensuring the system can generalize well to unseen data. By combining these pre-processing techniques, the model is equipped to handle medical images more effectively, leading to better segmentation outcomes.

## 3.3 Datasets Used

By combining these two diverse datasets BraTS 2020 for brain tumor segmentation and ISIC 2018 for skin lesion segmentation our model is exposed to a wide range of medical image segmentation challenges. This helps ensure that the model is not only effective in a specific domain but also generalizable across different medical contexts.

### 3.3.1 BraTS 2020

The BraTS challenge has become a leading benchmark for assessing the efficacy of advanced brain tumor segmentation techniques using multi-modal MRI data. We utilize the BraTS 2020 dataset [15] [16] [17] as a benchmark to evaluate the performance of our proposed model. This dataset consists of 3064 paired MRI brain images and their corresponding binary segmentation masks, which delineate the regions affected by brain tumors. The multi-modal MRI scans include four modalities for each subject:

*T1-weighted (T1) images:* Captures standard anatomical details.
*T1 post-contrast (T1Gd):* Highlights tumor regions via contrast-enhanced imaging.
*T2-weighted (T2):* Emphasizes the surrounding edema in brain tissue.
*Fluid-attenuated inversion recovery (FLAIR):* Suppresses the cerebrospinal fluid (CSF) signal to highlight hyperintense lesions.

Each brain MRI scan is accompanied by ground truth labels that divide the tumor into the following regions:
Whole Tumor (W.T.): Includes the entire tumor region.
Tumor Core (T.C.): Includes both necrotic and enhancing tumor parts.
Enhancing Tumor (E.T.): Consists of actively enhancing tumor regions visible in contrast-enhanced imaging.

This dataset presents unique challenges for segmentation due to the heterogeneous appearance of tumors, the high variability in tumor size and location, and the complex nature of MRI contrasts between different modalities. The BraTS 2020 dataset is well-suited for evaluating models aimed at robust and accurate segmentation of brain tumors.

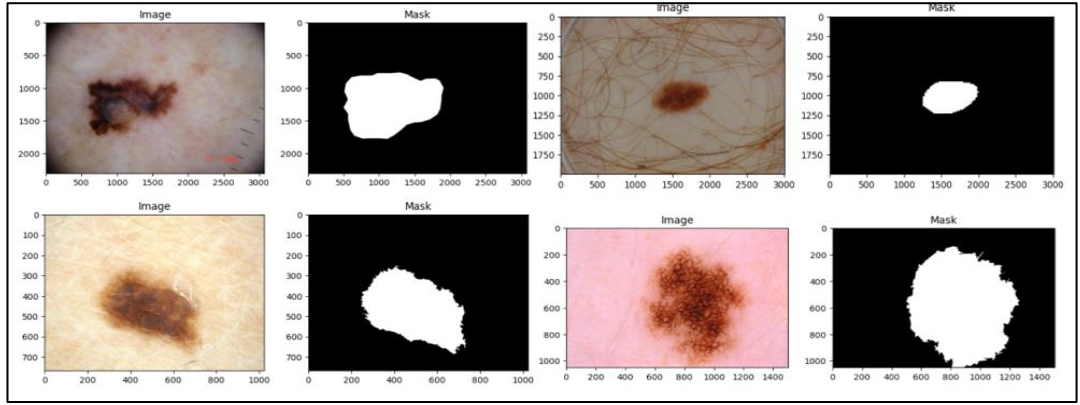

Figure 3:Sample images along with their corresponding mask from the ISIC 2018 dataset.

### 3.3.2 ISIC 2018

To enhance the generalization capabilities of our model, we incorporate data from the International Skin Imaging Collaboration (ISIC), a global initiative aimed at advancing melanoma diagnosis through datasets and challenges like the Skin Lesion Analysis Toward Melanoma Detection challenge. The ISIC 2018 dataset [20] consists of 2594 high-quality dermoscopic images, each paired with a segmentation mask that serves as the ground truth for skin lesion segmentation.

Automatic skin lesion segmentation presents unique challenges due to the diversity and variability in lesion characteristics. These include:

*Lesion size and shape variability:* Lesions exhibit significant differences in their dimensions and geometric forms, making uniform segmentation difficult.

*Color diversity:* Lesions appear in a wide range of hues, further complicated by varying pigmentation patterns within a single lesion.

*Occlusions by hair and other artifacts:* Dermoscopic images often contain hair and skin surface artifacts that occlude parts of the lesion, hindering the segmentation process.

*Low contrast:* Some lesions have poor contrast compared to the surrounding healthy skin, making their boundaries ambiguous and challenging to segment.

*Complex borders:* Lesion borders often lack clear definition, exhibiting irregular and fuzzy patterns that are difficult for both human annotators and automated systems to precisely delineate.

The presence of these challenges within the ISIC 2018 dataset makes it an ideal benchmark for evaluating the generalizability and robustness of skin lesion segmentation models. As shown in Figure 3, the variability in lesion characteristics requires sophisticated techniques to accurately segment the lesions from dermoscopic images.

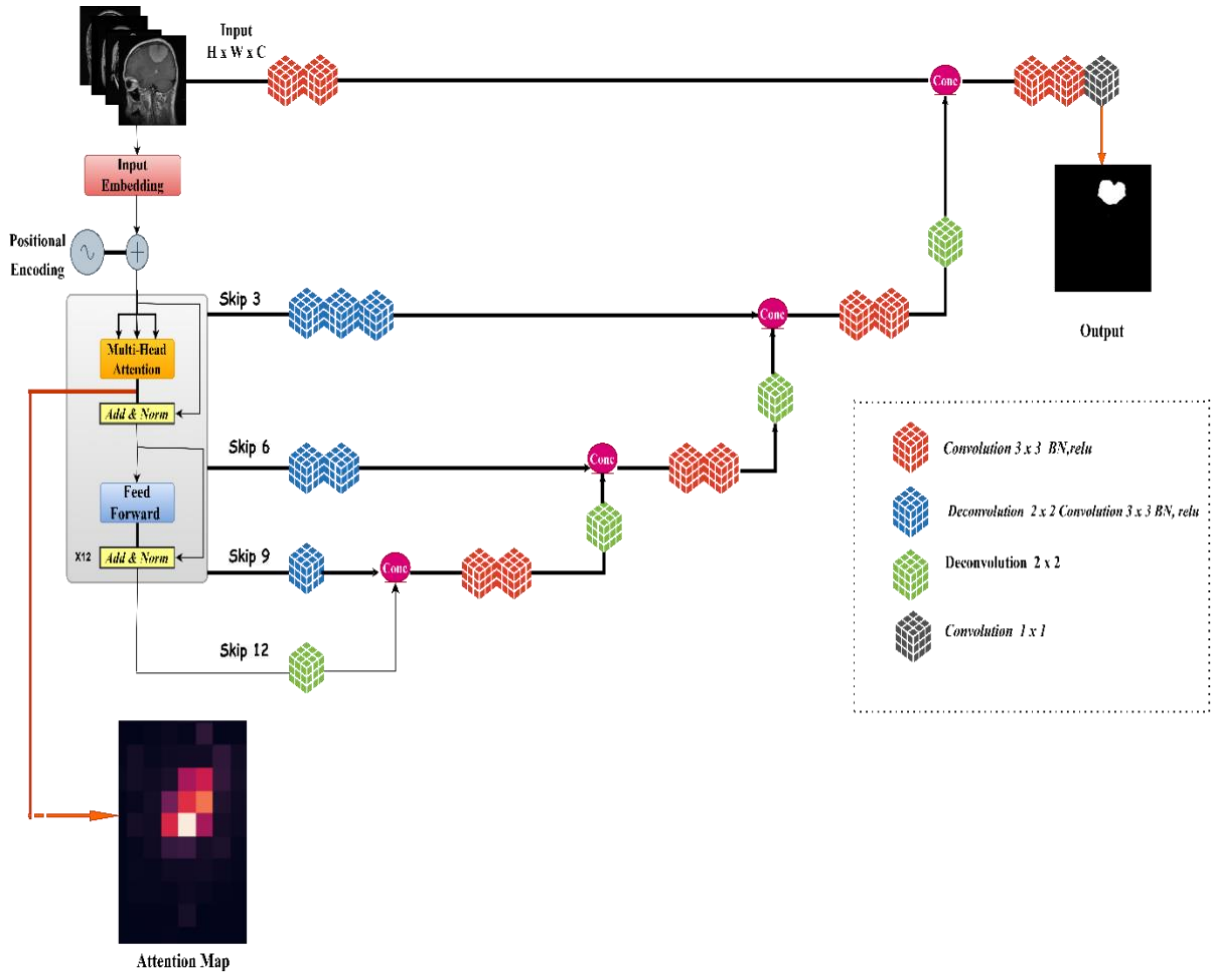

Figure 4: MAP-UNetR architecture overview. The input image is first augmented with positional embeddings and then processed through a Transformer model integrated within the MAP-UNETR framework. This architecture facilitates effective feature extraction and segmentation by leveraging both Transformer and U-Net components.

## 3.4 Proposed Transformer Architecture

The structure of the proposed MAPUNetR architecture, depicted in Figure 4 and detailed algorithm in Figure 5, adopts a bifurcated contracting expanding framework. This design leverages the strengths of both transformer modules for feature extraction and convolutional operations for accurate segmentation, striking a balance between global contextual understanding and local spatial precision.

The contracting pathway forms the encoder component of the model, consisting of a cascade of transformer blocks. Each block systematically extracts and refines hierarchical feature representations from the input data. The encoder focuses on learning both low- and high-level abstract features, capturing long-range dependencies by utilizing multi-head self-attention (MSA). The encoder progressively down samples the input, reducing its spatial resolution while enhancing the richness of the extracted features.

In parallel, the expanding pathway, functioning as the decoder, restores the spatial resolution of the feature maps through up sampling operations. Crucially, this expanding pathway is interconnected with the contracting pathway via skip connections. These skip connections allow direct information transfer between corresponding encoder and decoder stages, preserving fine-grained details that might otherwise be lost during the down sampling process in the encoder. This mechanism enables the model to effectively capture both the global context and the local intricacies, thus improving its capacity for detailed image analysis and reconstruction.

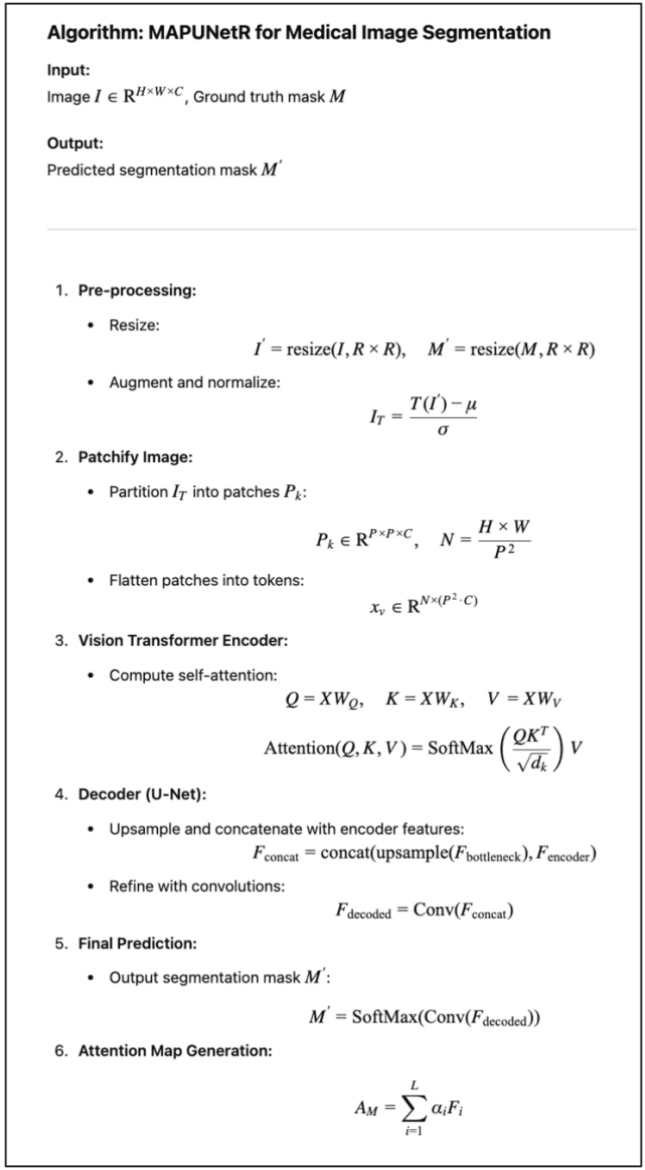

Figure 5: A step-by-step algorithmic approach to medical image segmentation using the proposed Map-UNETR model.

### 3.4.1 Transforming 2D Images into a 1D Sequence

In traditional natural language processing (NLP) tasks, transformers operate on 1D sequences. To adapt this paradigm for image data, the input image $x \in \mathbb{R}^{H \times W \times C}$, with height $H$, width $W$, and channel count $C$, is partitioned into non-overlapping patches of size $P \times P$. Each patch is then flattened and treated as an individual token in the sequence. The total number of patches $N$ is given by:

$$N = \frac{H \times W}{P^2} \quad (12)$$

Thus, the input image is transformed into a sequence of patches $x_v \in \mathbb{R}^{N \times (P^2 \cdot C)}$, where $N$ denotes the sequence length and $P^2 \cdot C$ represents the flattened size of each patch.

### 3.4.2 Multi-Head Self-Attention (MSA)

The MSA mechanism forms the core of the transformer. MSA comprises multiple self-attention layers operating in parallel. Each S.A. layer transforms the input sequence $Z \in \mathbb{R}^{N \times D}$ (where $N$ is the sequence length and $D$ is the embedding dimension) into a set of query (Q), key (K), and value (V) matrices. These matrices are calculated as:

$$Q = XW_Q, \ K = XW_K, \ V = XW_V \tag{13}$$

Here, $X$ is the input embedding matrix, $W_Q, W_K$ and $W_V$ are learnable weight matrices for the query, key, and value projections, respectively. The attention mechanism computes a weighted sum of the values, where the weights are determined by the similarity between queries and keys.

To stabilize the attention scores, the dot product between the queries and keys is normalized by the square root of the key vector's dimensionality, $\sqrt{d_k}$:

$$Attention(Q, K, V) = Softmax\left(\frac{QK^T}{\sqrt{d_k}}\right)V \tag{14}$$

This scaling ensures that the values in the attention matrix remain in a range conducive to stable gradients, particularly when the dimensionality $d_k$ is large. Without this scaling, the softmax function might drive the gradients towards zero, impairing the learning process.

### 3.4.3 Skip Connections and Decoder
The encoder's final output, commonly referred to as the bottleneck, is upsampled through deconvolution layers to increase its spatial dimensions. To enhance feature learning at higher resolutions, the upsampled feature maps are concatenated with their corresponding feature maps from the encoder via skip connections. This concatenation operation allows the decoder to recover fine-grained details while incorporating the broader context learned by the encoder. The concatenated feature maps are further processed by a sequence of $3 \times 3$ convolutional layers to refine the predictions. Each decoding step is followed by additional upsampling operations, progressively increasing the spatial resolution of the feature maps. The final layer of the decoder applies a $1 \times 1$ convolution, followed by a softmax activation, to generate the final segmentation mask.

### 3.4.4 Attention Maps for Interpretability
One of the key advantages of employing a vision transformer (ViT) encoder is its ability to provide attention maps, which offer insights into the model's decision-making process. These maps highlight the regions of the input image that the model focuses on during segmentation, providing interpretability. Attention maps visualize the areas of the image that contributed most to the model's final prediction, offering transparency into how the model understands and processes the input. The detailed algorithm is given in figure 6.

By leveraging attention mechanisms in the vision transformer, the proposed model can interpret its own predictions, enhancing explainability in critical medical image analysis tasks. This interpretability is particularly valuable in healthcare, where understanding the model's rationale is crucial for clinical decision-making.

## 4 Experiments and Results
The training process of the MAPUNetR model involved fine-tuning several key hyperparameters to ensure optimal performance and computational efficiency. Stochastic gradient Descent (SGD) was chosen as the optimizer for this task due to its robustness in large-scale image processing. The SGD optimizer was paired with the dice Loss function to minimize the error during training, maximizing the overlap between predicted and ground truth segmentation masks. A learning rate scheduler with a decay factor γ=0.1 was employed to ensure stable optimization throughout the training process. The learning rate started at an initial value of 0.01, and after every 20 epochs, the scheduler reduced the learning rate by a factor of 0.1. This dynamic adjustment allowed the model to make larger updates at the start of training and finer adjustments as it approached convergence, avoiding overshooting or getting stuck in local minima. After extensive experimentation, a batch size of 8 was selected for both the BraTS 2020 and ISIC 2018 datasets. This batch size provided a good balance between computational efficiency and the ability to fully utilize the available GPU memory. To improve the model's robustness and reduce overfitting, batch normalization was applied throughout the network, stabilizing the learning process by reducing internal covariate shifts. The model was trained for 70 epochs, a value determined through empirical testing to strike a balance between convergence and avoiding overfitting. The learning rate decay ensured that training continued to progress, while the number of epochs provided sufficient time for the model to reach its optimal configuration.

For our model, the total number of parameters are 2,398,193, of which 2,396,465 are trainable parameters, and 1,728 are non-trainable. The large proportion of trainable parameters (over 99.9%) reflects the model's high flexibility and its ability to learn complex features from the input data. The non-trainable parameters likely represent fixed elements such as batch normalization components, which do not require gradient updates during training but contribute to the model's stability and performance. This parameter distribution demonstrates the model's significant capacity for learning, indicating its suitability for tackling complex medical image segmentation tasks. This combination of hyperparameters and training strategy enabled the MAPUNetR model to achieve high segmentation accuracy while maintaining strong generalization capabilities across both datasets.

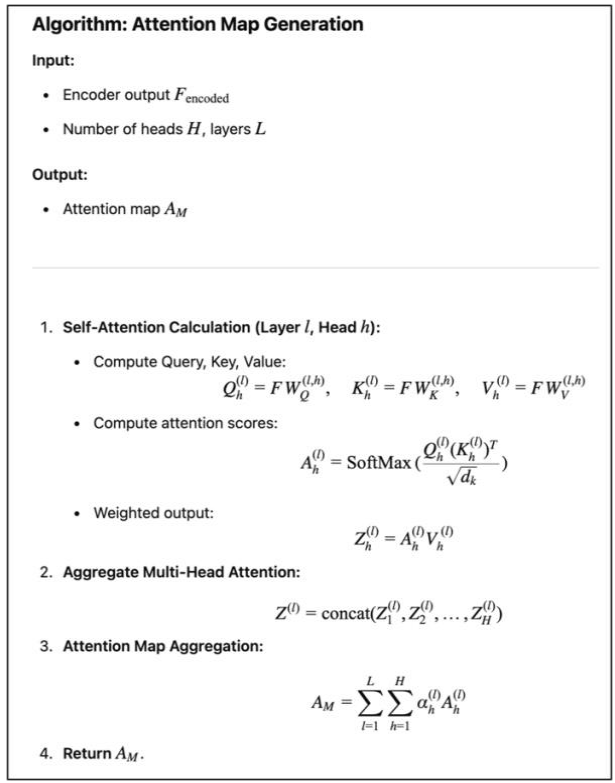

Figure 6: This figure illustrates the steps involved in generating attention maps from the Vision Transformer encoder.

### 4.1 Evaluation Metrics:

The efficacy of our segmentation model was assessed using a suite of widely recognized computer vision metrics. These include Dice Coefficient (DSC), Intersection Over Union (IoU), Overall Accuracy (Acc), Precision (p), Recall (r),

In addition to these performance metrics, in parallel, we assessed the efficacy of the dice loss as the objective loss function for our model. The mathematical formulations for these metrics are as follows

a) *Dice coefficient :* This numerical index quantifies the extent of concordance between the algorithm-generated and reference segmentation masks, offering a meaningful gauge of segmentation precision.

$$\text{DSC} = 2 \times \frac{\text{Area of Overlap}}{\text{Total area of masks}} = 2 \times \frac{|P \cap G|}{|P \cup G|} \quad (15)$$

Where the variable P denotes the predicted set and G is the ground truth set.

b) *Intersection Over Union (IoU):* Alternatively known as the Jaccard coefficient, this measure evaluates the congruence between the computational segmentation result and the validated reference standard. It expresses the relationship between the overlapping region and the combined area of the predicted and actual segmented domains

$$\text{IoU} = \frac{\text{Area of Overlap}}{\text{Total area of masks}} = \frac{|P \cap G|}{|P \cup G|} \quad (16)$$

c) *Accuracy (A):* This metric evaluates the model's comprehensive accuracy by determining the proportion of accurate classifications (encompassing both correctly identified positives and negatives) relative to the total instances analyzed. It offers a broad appraisal of the model's efficacy across all categories, reflecting its capacity to accurately categorize both affirmative and negative cases within the complete dataset

$$A = \frac{True_{positive} + True_{negative}}{True_{postive} + True_{negative} + False_{postive} + False_{negative}} \quad (17)$$

d) *Precision(p):* Precision gauges the model's proficiency in correctly identifying positive cases within the subset of instances it classifies as positive. This metric is calculated by taking the number of correctly identified positive cases and dividing it by the total positive results, encompassing both right and wrong classifications.

$$p = \frac{True_{positive}}{True_{positive} + False_{positive}} \quad (18)$$

e) *Recall (r):* It quantifies the model's ability to detect all relevant cases. It is calculated as the ratio of correctly identified positive instances (true positives) to the total actual positive instances (true positives plus false negatives) in the dataset. This metric assesses the model's sensitivity in capturing all pertinent cases, indicating its effectiveness in minimizing false negatives.

$$r = \frac{True_{positive}}{True_{positive} + False_{negative}} \quad (19)$$

## 4.2 Quantitative and Qualitative results

This section presents the evaluation of MAPUNetR using both quantitative metrics, such as dice coefficient and IoU, and qualitative visualizations of segmentation outputs. Together, these results provide a clear understanding of the model's accuracy, performance, and interpretability in medical image segmentation.

### 4.2.1 Results on BraTS 2020

In this section, we rigorously evaluate the performance of the proposed MAPUNetR model on the BraTS 2020 dataset, which comprises multi-modal MRI scans for brain tumour segmentation. The complexity of the BraTS 2020 dataset, which includes varied tumour sub-regions such as the whole tumour (W.T.), tumour core (T.C.), and enhancing tumour (E.T.), poses significant challenges for segmentation algorithms due to the heterogeneity in tumour appearance and the presence of noise in the medical images. Despite these challenges, the proposed model achieved a dice similarity coefficient of 0.88 within 100 epochs, demonstrating its capacity to accurately capture tumour regions with minimal training time compared to other state-of-the-art methods.

### 4.2.2 Model Performance Comparison:

Table 1 presents a comparative analysis of the dice score achieved by our model in relation to other leading segmentation architectures on the BraTS 2020 benchmark. Models such as BiTr-Unet, Swin UNETR, and TransBTS required significantly longer training periods to achieve their respective dice coefficients. For instance, BiTr-Unet achieved a dice score of 0.8187 after 7050 epochs, while Swin UNETR, a transformer-based model, reported a dice score of 0.871 after 800 epochs. Notably, MAPUNetR achieved a superior dice score of 0.88 after only 100 epochs, highlighting the model's efficiency and ability to converge rapidly with fewer iterations, reducing computational overhead and training time.

The performance superiority of MAPUNetR can be attributed to the hybrid architecture that synergizes the global feature extraction capabilities of the vision transformer (ViT) encoder with the detailed, spatially aware decoder of the U-Net framework. By leveraging attention mechanisms and channel-spatial refinement, our model efficiently processes large-scale medical image datasets while preserving fine-grained details crucial for accurate segmentation.

Table 1: The table presents the Dice Similarity Coefficient (DSC) and the number of epochs used for various models. The proposed model demonstrates improved DSC compared to other methods.

| Dataset | Method | Author | DSC | Epochs |
|---|---|---|---|---|
| BRATS | BiTr-Unet[21] | Qiran Jia and Hai Shu | 0.8187 | 7050 |
| BRATS | Swin UNETR[22] | Hatamizadeh et al | 0.871 | 800 |
| BRATS | TransBTS[23] | Wang et al | 0.7873 | 8000 |
| BRATS | nnformer[24] | Zhou et al | 0.86 | 1000 |
| BRATS | UNETR[25] | Hatamizadeh et al | 0.84 | 20000 |
| BRATS | MISSU[26] | Wang et al | 0.80 | 1000 |
| **BRATS** | **Proposed Model** | | **0.88** | **100** |

### 4.2.3 Epoch-wise Performance Analysis:

To better understand the learning dynamics of MAPUNetR, we tracked the performance metrics across the first five and last five epochs of the training process, as shown in Tables 2 and 3.

Table 2: Highlight the model's Metrics of first 5 epochs

| Epoch | Accuracy | Dice_coef | Loss | L. Rate | val_acc | val_dice_coef | val_loss |
|---|---|---|---|---|---|---|---|
| 0 | 0.817330 | 0.181723 | 0.818314 | 0.1 | 0.832176 | 0.159865 | 0.840050 |
| 1 | 0.973131 | 0.380915 | 0.619105 | 0.1 | 0.970950 | 0.422430 | 0.577503 |
| 2 | 0.976018 | 0.461093 | 0.538944 | 0.1 | 0.977660 | 0.510241 | 0.489708 |
| 3 | 0.978396 | 0.523247 | 0.476756 | 0.1 | 0.975699 | 0.533661 | 0.466174 |
| 4 | 0.489708 | 0.557330 | 0.442682 | 0.1 | 0.979424 | 0.572978 | 0.426897 |

Table 3: Highlight the model's Metrics of last 5 epochs.

| Epoch | Accuracy | Dice_coef | Loss | L. Rate | val_acc | val_dice_coef | val_loss |
|---|---|---|---|---|---|---|---|
| 64 | 0.9941771 | 0.8870785 | 0.112919 | 9.999999e-06 | 0.986976 | 0.71764135360 | 0.2823975682 |
| 65 | 0.9941831 | 0.8871771 | 0.112820 | 9.999999e-06 | 0.986974 | 0.71774232387 | 0.2822965681 |
| 66 | 0.9941951 | 0.8873092 | 0.112687 | 9.999999e-06 | 0.986975 | 0.71766471862 | 0.2823740839 |
| 67 | 0.9941736 | 0.8869324 | 0.113067 | 9.999999e-06 | 0.986974 | 0.71771687269 | 0.2823221087 |
| 68 | 0.9941851 | 0.8871397 | 0.112860 | 9.999999e-06 | 0.986974 | 0.71773564815 | 0.2823032140 |

In the early stages, as seen in Table 2, the dice coefficient improved significantly, rising from 0.1817 at epoch 0 to 0.5573 at epoch 4. The model demonstrated rapid learning, which is further substantiated by the increase in accuracy and the corresponding decrease in loss. By the final epochs (Table 3), the dice coefficient stabilized around 0.8871, and both training and validation losses plateaued, indicating that the model had converged. These results show that the model efficiently captures the spatial hierarchies and feature relationships necessary for accurate segmentation within a limited number of training iterations.

### 4.2.4 Learning Progression and Generalization:

The learning progression of the MAPUNetR model was further analysed through the plots of accuracy, loss, and dice coefficient over training and validation datasets, as depicted in Figures 7(a), 7(b), and 7(c).

Figure 7(a) shows the accuracy curves for both training and validation sets, which consistently improve over the course of training. The convergence of the validation accuracy towards the training accuracy reflects the model's strong generalization capabilities, as it performs well not only on the training data but also on unseen validation samples. Figure 7(b) illustrates the loss curves for both datasets. The consistent reduction in loss indicates that the model is steadily minimizing the prediction error between the generated segmentation mask and the ground truth. The smooth downward trajectory of both training and validation loss suggests that overfitting is minimal, due in part to the regularization mechanisms like batch normalization applied throughout the network. Figure 7(c) presents the dice coefficient curves, which are a direct measure of segmentation quality. The training and validation dice coefficients rise in tandem, with the validation dice approaching the training dice coefficient as training progresses. This further supports the conclusion that the model is generalizing well across datasets and has learned to effectively delineate tumour boundaries.

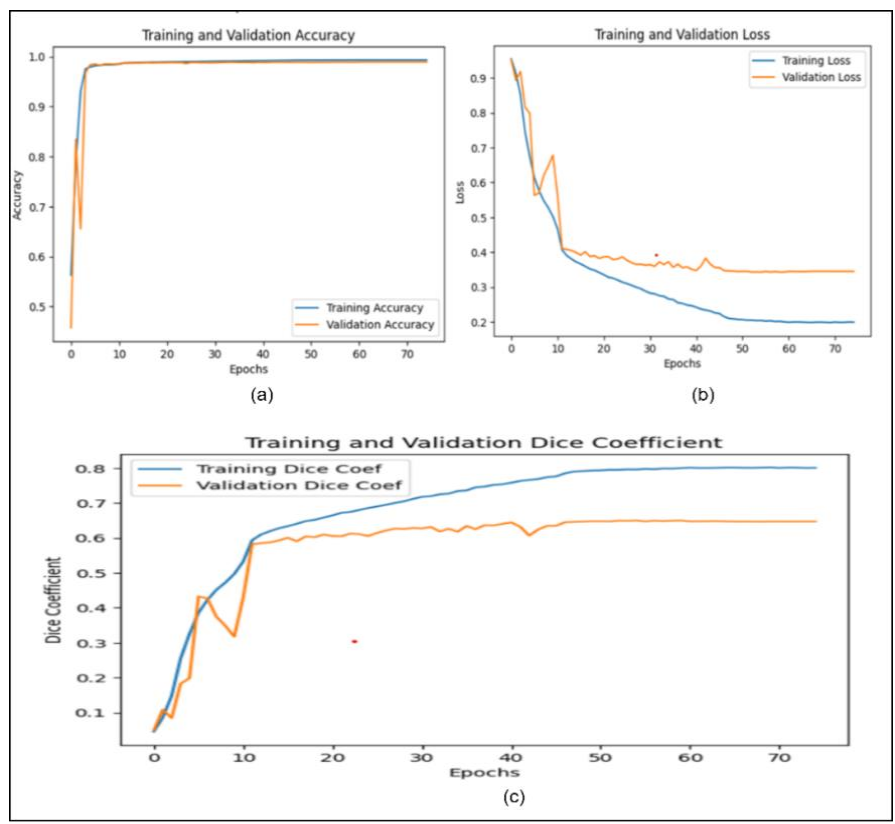

Figure 7:(a) Depicts the evolution of model accuracy during training and validation on the BraTS dataset. (b) Depicts the training and validation loss curves for the BraTS dataset.(c) Dice coefficient curve.

### 4.2.5 Qualitative Analysis and Visualization:

In addition to the quantitative evaluation, qualitative results were visualized to further assess the performance of the MAPUNetR model. Figure 8 showcases a comparison between the segmentation mask produced by the model and the ground truth mask for a representative sample from the BraTS 2020 dataset. The predicted mask closely aligns with the ground truth, with precise delineation of the tumour boundary, indicating that the model is adept at capturing both the global context and local spatial details critical for accurate tumour segmentation.

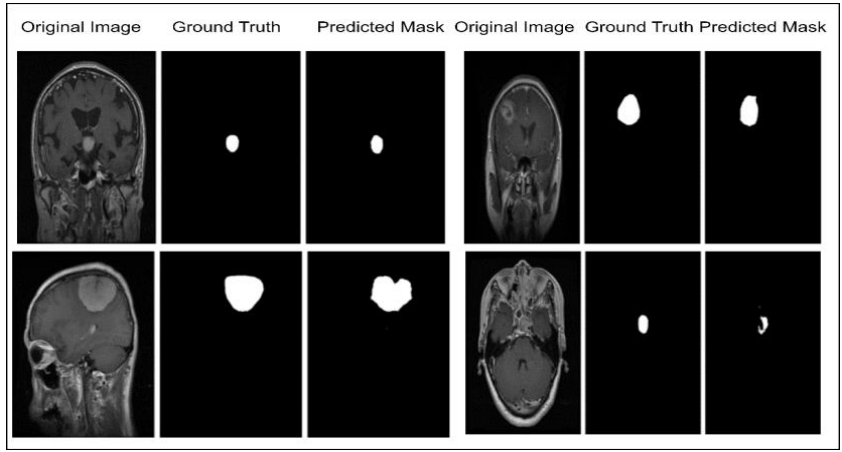

Figure 8: Illustration of the segmentation mask produced by the model, demonstrating the delineation of regions identified within the medical image based on the model's predictions.

### 4.2.6 Attention Map Visualization:

To enhance interpretability, attention maps derived from the encoder's multi-head attention layers were generated, providing insights into the regions of the input image that the model focuses on during prediction. Figure 9 illustrates these attention maps, highlighting the area's most influential in the model's decision-making process. The attention maps confirm that the model is prioritizing relevant regions within the MRI scans, such as tumour boundaries and core areas, which are crucial for accurate segmentation.

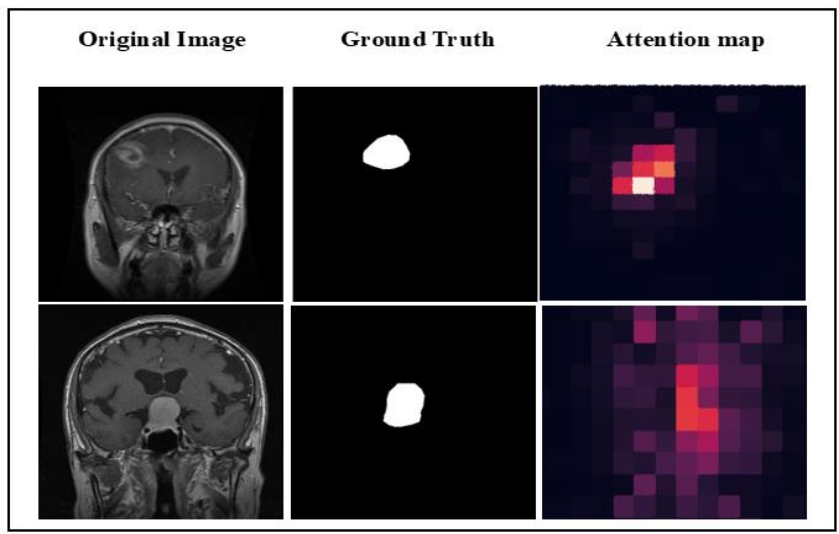

Figure 9: Visualization of attention maps that indicate the regions of the image which contribute the most to the model's prediction, offering insights into the model's decision-making process.

Overall, the MAPUNetR model shows exceptional performance on the BraTS 2020 dataset, achieving high dice coefficients with a fraction of the training time compared to other state-of-the-art models. The quantitative results, coupled with qualitative analyses and interpretability through attention maps, highlight the model's ability to perform accurate and reliable brain tumour segmentation. The model's efficiency, stability, and generalization capabilities make it a valuable tool for medical image analysis, particularly in the domain of neuroimaging.

### 4.3 Results on ISIC 2018

This section discusses the performance of the proposed MAPUNetR model on the ISIC 2018 skin lesion segmentation dataset. The ISIC 2018 dataset introduces additional challenges due to the variability in lesion size, shape, and contrast, making it a benchmark for evaluating segmentation models in dermatological image analysis.

Our model achieved a dice coefficient of 0.927 after 70 epochs of training, demonstrating its superior capability to handle such variability, and it outperformed several state-of-the-art models, as presented in Table 4.

### 4.3.1 Model Performance Comparison:

Table 4 provides a comparison between MAPUNetR and other models applied to the ISIC 2018 dataset. Notably, MAPUNetR surpassed models such as MT-TransUNet and DS-TransUNet, which required significantly more training epochs. For example, MT-TransUNet achieved a dice score of 0.800 after 40,000 epochs, whereas MAPUNetR reached a dice score of 0.927 after just 70 epochs. This efficiency can be attributed to the attention-driven feature extraction capability of the vision transformer encoder, which allows the model to capture both global context and fine-grained local details in the images.

This performance demonstrates the efficiency of the hybrid vision transformer and U-Net architecture. The attention mechanism in the transformer enables efficient feature extraction across various scales, which is crucial for handling the highly variable appearance of skin lesions in the dataset.

Table 4: Comparison of different methods applied to the skin lesion dataset, highlighting the Dice Similarity Coefficient and the number of epochs used in the training process. The proposed model demonstrates superior performance.

| Dataset | Method | Author | DSC | EPOCHS |
|---|---|---|---|---|
| SKIN | MT-TransUNet[27] | Chen et al | 0.800 | 40000 |
| SKIN | DS-TransUNet [28] | Lin et al | 0.868 | 800 |
| SKIN | BEFUnet[10] | Manzari et al | 0.868 | N.A |
| SKIN | TransUNet[29] | Chen et al | 0.849 | N.A |
| SKIN | TransNorm[30] | Azad et al | 0.895 | 100 |
| SKIN | Swin-Unet[31] | Cao et al | 0.894 | N.A. |
| SKIN | BAT[32] | Wang et al | 0:912 | 500 |
| SKIN | HiFormer[7] | Heidari et al | 0.910 | N.A |
| SKIN | MedT[33] | Valanarasu et al | 0.859 | N.A. |
| SKIN | UNet++[34] | Zhou et al | 0.879 | N.A |
| SKIN | SBE[35] | Lee et al | 0.918 | 4000 |
| **SKIN** | **Proposed Model** | | **0.927** | **70** |

### 4.3.2 Epoch-wise Performance Analysis:

Tables 5 and 6 provide detailed performance metrics from the initial and final stages of training. The rapid rise in the dice coefficient during the first five epochs (Table 5) highlights the model's ability to learn effectively from the start, reaching a dice coefficient of 0.823 by epoch 4. The final five epochs (Table 6) reflect the model's stabilization at a high dice coefficient of 0.927, demonstrating consistent performance across training and validation data.

Table 5: Performance metrics for the first five epochs of training on the ISIC 2018 dataset. Metrics include accuracy, Dice coefficient, loss, learning rate, validation accuracy, validation Dice coefficient, and validation loss.

| Epoch | Accuracy | Dice_coef | Loss | L. Rate | val_acc | val_dice_coef | val_loss |
|---|---|---|---|---|---|---|---|
| 0 | 0.830806851 | 0.7225411 | 0.277477 | 0.1 | 0.360044 | 0.4739192128 | 0.52677947 |
| 1 | 0.868051648 | 0.7802910 | 0.219636 | 0.1 | 0.802094 | 0.5399222373 | 0.46115270 |
| 2 | 0.878080427 | 0.7973234 | 0.202587 | 0.1 | 0.848831 | 0.7527146339 | 0.24790471 |
| 3 | 0.888683140 | 0.8138136 | 0.186091 | 0.1 | 0.883516 | 0.7808754444 | 0.2196629 |
| 4 | 0.894138038 | 0.8234840 | 0.176449 | 0.1 | 0.744325 | 0.2752812206 | 0.72567123 |

Table 6: Summary of the performance metrics for the last five epochs of training on the ISIC 2018 dataset.

| Epoch | Accuracy | Dice_coef | Loss | L. Rate | val_acc | val_dice_coef | val_loss |
|---|---|---|---|---|---|---|---|
| 65 | 0.95564967 | 0.92713344 | 0.072878 | 9.999999e-07 | 0.945518 | 0.904070198535 | 0.096156 |
| 66 | 0.95573157 | 0.92727959 | 0.072731 | 9.999999e-07 | 0.945516 | 0.904070794582 | 0.096156 |
| 67 | 0.95570331 | 0.92732656 | 0.072685 | 9.999999e-07 | 0.945502 | 0.904063999652 | 0.0961634 |
| 68 | 0.95569413 | 0.92728292 | 0.072728 | 9.999999e-07 | 0.945503 | 0.904057979583 | 0.0961693 |
| 69 | 0.95571088 | 0.92731744 | 0.072696 | 9.999999e-07 | 0.945491 | 0.904046535491 | 0.0961807 |

By epoch 69, the model achieves near-optimal performance, with the dice coefficient stabilizing around 0.9273 and validation metrics closely tracking the training metrics. This indicates the model's robustness and its ability to generalize well across unseen data.

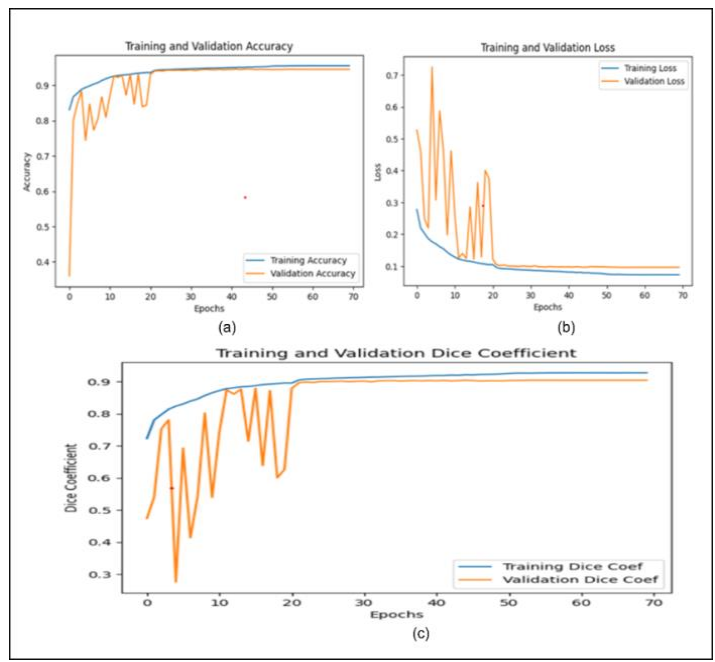

Figure 10: a) Performance trajectory depicting training and validation set precision on the ISIC 2018 dataset, b) depicts the training and validation loss curves for the ISIC 2018. C) Dice coefficient Curve.

### 4.3.3 Learning Progression:

The learning progression of the MAPUNetR model is further visualized through accuracy, loss, and Dice coefficient plots for both training and validation sets, shown in Figure 10.

Figure 10(a) shows that both training and validation accuracy steadily improve throughout the training process, indicating strong model generalization. Figure 10(b) illustrates the smooth decrease in loss for both training and validation, confirming the model's ability to reduce prediction error consistently. Figure 10(c) highlights the increase in dice coefficient, reflecting improvements in segmentation quality over time. The convergence between training and validation dice coefficients demonstrates minimal overfitting, with the model learning the essential patterns required for accurate segmentation.

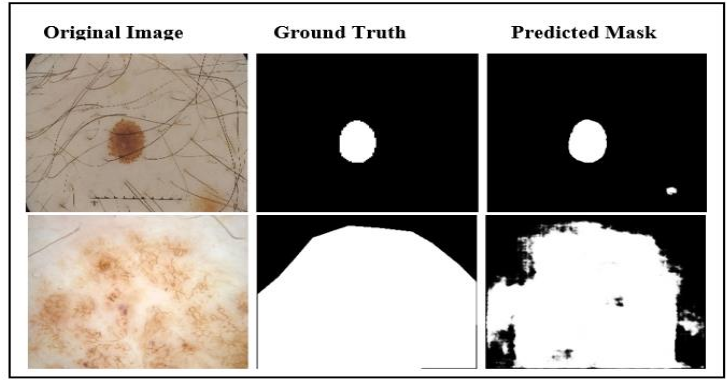

Figure 11: *Segmentation results on the ISIC 2018 Dataset.* *The figure showcases the segmentation performance of the proposed model on various images from the ISIC 2018 dataset. The segmentation masks are overlaid on the original images to illustrate the accuracy and effectiveness of the model in delineating skin lesions.*

### 4.3.4 Qualitative Analysis:
Qualitative results on the ISIC 2018 dataset further substantiate the quantitative findings. As shown in Figure 11, the segmentation masks generated by MAPUNetR closely align with expert-annotated ground truth masks. These results indicate that the model accurately delineates lesion boundaries, effectively handling the challenges posed by lesion variability in size, shape, and contrast.

### 4.3.5 Attention Map Visualization:
Figure 12 visualizes the attention maps generated by the vision transformer encoder. These maps provide insight into the regions of the input images that contribute most to the model's predictions. The attention maps reveal that the model focuses on critical areas of the skin lesions, ensuring accurate segmentation even in complex cases. This interpretability is vital for clinical applications, as it allows practitioners to understand and trust the model's decision-making process.

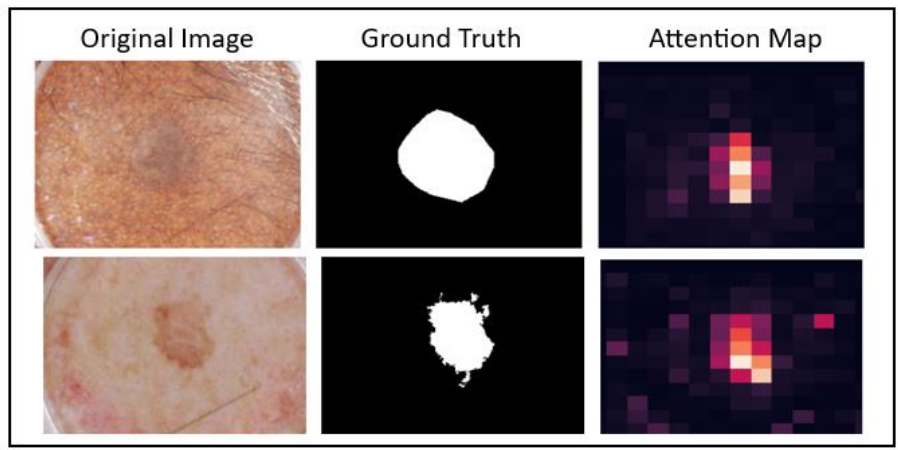

Figure 12: **Generated attention maps for the ISIC Dataset.** The figure displays the attention maps produced by the proposed model, highlighting the regions of the images from the ISIC dataset that have the most influence on the model's predictions. These attention maps provide insight into which parts of the images are critical for accurate segmentation.

In conclusion, the MAPUNetR model demonstrates superior performance on the ISIC 2018 dataset, with a dice coefficient of 0.927. The model's efficient use of attention mechanisms, combined with the U-Net decoder's ability to capture fine details, enables it to outperform competing models. The strong quantitative and qualitative results, along with the interpretability provided by attention maps, position MAPUNetR as a robust tool for dermatological image segmentation, with potential applications in automated skin lesion analysis.

## 5. Conclusion and Future work
In this study, we presented MAPUNetR, a hybrid architecture combining vision transformers and U-Net for medical image segmentation. The model achieved dice coefficients of 0.88 on the BraTS 2020 dataset and 0.927 on the ISIC 2018 dataset, outperforming several state-of-the-art models with fewer training epochs. The vision transformer's self-attention mechanism enabled efficient global context understanding, while the U-Net decoder preserved essential spatial details, resulting in accurate and interpretable segmentation outputs. The model's attention maps further enhanced interpretability, providing clinicians with insights into the decision-making process, which is crucial for clinical adoption. Future work will focus on addressing key limitations, such as the availability of larger and more diverse datasets to improve model generalizability. Collaborations with multiple healthcare institutions will help amass such datasets. Additionally, leveraging more powerful hardware, like high-RAM GPUs, will enhance the training process, allowing for more comprehensive model development. Further integration of advanced explainability methods will improve model transparency and interpretability, fostering trust in AI-assisted diagnostic tools. These efforts aim to create a more robust and widely applicable medical image segmentation model for use in clinical settings.

## Declarations

**Conflict of Interest**: The authors declare no conflict of interest.

**Ethical approval:** Not applicable.

**Funding:** Not applicable.

**Data availability:** We do not analyze or generate datasets because our work proceeds with a theoretical and mathematical approach. One can obtain the relevant materials from the references below.

**Author's contributions: Ovais Iqbal Shah:** Conceptualization, Methodology, Data curation, Writing- Original draft, Formal analysis, **Danish *Raza Rizvi*:** Supervision, Formal analysis, Writing- Reviewing and Editing, **Aqib Nazir Mir:** Conceptualization, Methodology, Data curation, Writing- Original draft, Visualization, Formal analysis.

**Human and animal participants**: Not applicable.